\documentstyle[12pt,epsfig]{article}
\begin{document}
\baselineskip 24pt
\begin{flushright} SNUTP 97-100 \\
\end{flushright}                     
\begin{center}
{\large \bf Soliton and Domain Wall in the Self-Dual 
$CP(1)$ Model}    
\vspace{.5cm}
Sung-Soo Kim $^{a,1}$, Phillial Oh $^{a,2}$, 
and Chaiho Rim $^{b,3}$ \\
$^a$ {\it Department of Physics, Sung Kyun Kwan University,
Suwon 440-746,  Korea }\\
$^b$ {\it Department of Physics, Chonbuk National University,
Chonju 561-756,  Korea }\\
\end{center}

\vspace{0.5cm}
\begin{center}
{\bf Abstract} \\
\end{center}
We perform  the dimensional reduction of the nonrelativistic
$CP(1)$ model coupled to an Abelian Chern-Simons gauge 
field in the self-dual limit, and investigate the soliton 
and domain wall solutions of the emerging 1+1 dimensional 
self-dual spin system. This system is described by
inhomogeneous Landau-Lifshitz system with an extra non-local 
term. The Hamiltonian is Bogomol'nyi bounded from below and has 
four adjusting parameters. The Bogomol'nyi equation is described 
in detail in analogy with the Newtonian equation of motion and 
its numerical solution is presented.

\hspace{0.3cm}

\noindent PACS codes: 11.10.Lm \\
\noindent Keywords: Self-dual soliton, Inhomogeneous
Landau-Lifshitz equation, Dimensional reduction, 
Chern-Simon gauge field, Bogomol'nyi bound, Domain wall\\

\noindent
\hbox to 10cm{\hrulefill}

\baselineskip 12pt

\noindent $^1$ {\small Electronic-mail: billy@newton.skku.ac.kr} \\
$^2$ {\small Electronic-mail: ploh@newton.skku.ac.kr}\\
$^3$ {\small Electronic-mail: rim@phy0.chonbuk.ac.kr}

\pagebreak

\baselineskip 24pt

\section{Introduction}
The recent investigation 
of the dimensional reduction of
the Chern-Simons-matter system in 
2+1 dimension \cite{jack-pi} and its modification 
uncovered many new features in the soliton physics of
1+1 dimension \cite{rabe,jack,min,kao,sen,ohrim} through
the discovery of novel soliton solutions \cite{jack}.
In particular, the reduction in the self-dual limit is interesting 
because it can lead to a lineal system which is Bogomol'nyi bounded 
and therefore, its ground state can be investigated with relative ease.
This possibility is demonstrated in Ref. \cite{ohrim}, where  
a self-dual formulation of the non-linear
Schr\"odinger equation (NLSE) in 1+1 dimension is obtained
through the dimensional reduction
of the Abelian self-dual Chern-Simons theory. 
The conventional second-order differential equation is replaced
by the first-order (non-local)  Bogomol'nyi equation 
which is equivalent to the completely integrable Liouville equation.
In the non-Abelian case, similar thing happens  
but with Toda equation \cite{jak}.
 
In this paper, we carry out the dimensional reduction of the
self-dual Chern-Simons non-relativistic spin system \cite{ohpark}
and study its solutions which saturate the Bogomol'nyi bound.
The model in consideration   
is  the non-relativistic $CP(1)$ model minimally coupled to an
Abelian Chern-Simons gauge field in a uniform background,
which  exhibits  a rich structure of rotationally 
symmetric self-dual Bogomol'nyi solitons depending on 
the  various background charges \cite{ohkim}.
The dimensionally reduced
model turns out to be the self-dual spin system
where  the inhomogeneous Landau-Lifshitz equation (ILLE) \cite{fadd}
is modified by a non-local interaction. 
Unlike the self-dual formulation of
NLSE \cite{ohrim,andr},
this non-local term can neither be replaced by 
a constant of motion nor 
be eliminated by a phase redefinition.
Rather,  it  is responsible for a variety of domain walls and 
magnetic solitons which are not  found in the usual 
spin system \cite{kose}.

The paper is organized as follows: 
In section 2, we perform the dimensional reduction of the 
2+1 dimensional Chern-Simon spin system 
to obtain the  lineal self-dual non-local
ferromagnet model. 
In section 3, its Bogomol'nyi bound is investigated
which leads to  the non-local Bogomol'nyi equation.
We discuss the features of  
solitons and domain walls 
with careful consideration on the range of parameters.
Their  numerical results are also  presented. 
Section 4 is the conclusion.\\

\section{Dimensional Reduction} 
In this section, we perform 
the dimensional reduction of the non-relativistic
self-dual Chern-Simons 
spin system \cite{ohpark,ohkim}
and obtain our lineal model. Let us consider the Lagrangian
\begin{eqnarray}
\cal{L} &=& i[(\Psi^\dagger\nabla_0\Psi)-(\nabla_0\Psi)^\dagger
            \Psi]+a_0(\Psi^\dagger\Psi -1)-\rho_eA_0\nonumber\\
        & & -2|D_i\Psi|^2-V_{[2]}(\Psi)+\frac{\kappa}{2}\epsilon^{\mu\nu\rho}
            A_\mu\partial_\nu A_\rho,\label{lagra}
\end{eqnarray}
where 
       $$ \nabla_\mu = \partial_\mu +
              i A_\mu\frac{\sigma^3}{2},\;\; 
              D_\mu=\nabla_\mu+ia_\mu.$$
$\Psi$ is a two component complex field,
$\Psi^\dagger=
            (\Psi^*_1\,,{}\Psi^*_2)$,
satisfying the $CP(1)$ constraint,
$\Psi^\dagger\Psi=1$,
which is imposed through the zeroth component of 
the auxilliary $U(1)$ gauge field
$a_\mu$.
$A_\mu$ is the $U(1)$ Chern-Simons gauge field,
$\rho_e$ is a uniform background charge,
and $V_{[2]}(\Psi)$ is a potential energy term
given below Eq.~(\ref{poten}). 
The above Lagrangian in $CP(1)$ representation 
is another expression of 
the one proposed
in Ref. \cite{ohpark,ohkim} where the Lagrangian is 
given in the coadjoint representation.
The first two terms in the Lagrangian are the
familiar canonical symplectic term which can be rewritten
as $2{\rm Tr}(Kg^{-1}\nabla_0 g),~K=i\sigma^3/2, g\in SU(2)$. 
When  the auxiliary field $a_i$ is eliminated 
with the help of the $CP(1)$ constraint, 
the Lagrangian is expressed in terms of the coadjoint orbit
variable, $Q=i\Psi\Psi^\dagger-iI/2$.
$CP(1)$ representation  is used here since it 
is more convenient for our purpose.  

Let us first construct the Bogomol'nyi bound. Using the identity
\begin{eqnarray}
(D_i\Psi)^\dagger(D_i\Psi)=&\left|
 (D_1\pm iD_2)\Psi\right|^2\pm\epsilon_{ij}
\partial_i\left((\Psi^\dagger\frac{\sigma^3}{2}
\Psi- w)A_j\right)\nonumber\\ 
&\mp i
\epsilon_{ij}\partial_i({\Psi}^\dagger\partial_j\Psi)
\mp\epsilon_{ij}(\partial_iA_j)
(\Psi^\dagger\frac{\sigma^3}{2}\Psi-w),\label{ident}
\end{eqnarray}
where $\omega$ is a free paameter, we find the Hamiltonian  to be
\begin{eqnarray}
H_{[2]}
&=&\int d^2x\left(2\left|D_i{\Psi}\right|^2+V(\Psi)_{[2]}\right)\nonumber\\
 &=&\int d^2x\left\{2\left|(D_1\pm iD_2)\Psi\right|^2
+V_{[2]} (\Psi)\mp 
    \frac{1}{2}F_{ij}(\Psi^\dagger\sigma^3\Psi-2w)\right\}
    \mp 4\pi T_{[2]}. \label{hamil}
\end{eqnarray}
$F_{ij}$ is the field strength of the gauge field and the topological 
charge is given by
\begin{equation}
T_{[2]} =\frac{1}{4\pi}\int d^2x\left\{2i\epsilon_{ij}\partial_i
(\Psi^\dagger\partial_j\Psi)+\epsilon_{ij}\partial_i\left[
(2w-\Psi^\dagger\sigma^3\Psi)A_j\right]\right\}.\label{charg}
\end{equation}
Replacing the gauge field strength  in terms of matter fields
with the help of  the Gauss's law constraint
\begin{equation}
\frac{\kappa}{2}\epsilon_{ij} 
F_{ij}=\Psi^\dagger\sigma^3\Psi + \rho_e,
\label{gauss}
\end{equation}
and choosing the form of the potential energy to be
\begin{equation}
V_{[2]}(\Psi)=\pm\frac{1}{\kappa}(Q^3+\rho_e)(Q^3-w),
~~Q^3\equiv\Psi^\dagger\sigma^3\Psi ,
\label{poten}
\end{equation}
we have  the Hamiltonian with the Bogomol'nyi bound 
\cite{ohpark,ohkim}
\begin{equation}
H_{[2]}=2\int d^2x\left|\left\{\left(\nabla_1-
\frac1{2}\Psi^\dagger
\nabla^{\hspace{-3.8mm}\raisebox{1.2mm}{$\leftrightarrow$}}_1
\Psi 
\right)
\pm i\left(\nabla_2-
\frac1{2}\Psi^\dagger
\nabla^{\hspace{-3.8mm}\raisebox{1.2mm}{$\leftrightarrow$}}_2
\Psi 
\right)\right\}
\Psi\right|^2\mp4\pi T_{[2]},
~~\Psi^\dagger\Psi=1,\label{bolimit}
\end{equation}
where $a_i$ has been eliminated by using the equation of
motion:
\begin{equation}
a_i=\frac{i}{2}\Psi^\dagger
\partial^{\hspace{-2.9mm}\raisebox{1.2mm}{$\leftrightarrow$}}_i\Psi - 
    \Psi^\dagger\frac{\sigma^3}{2}\Psi A_i = 
\frac{i}{2}\Psi^\dagger\nabla^{\hspace{-3.8mm}\raisebox{1.2mm}{$\leftrightarrow$}}
_i\Psi.\label{eqofm}
\end{equation}

To obtain  the Hamiltonian which is dimensionally
reduced to the one spatial dimension, 
we regard  $\Psi$ to be  independent of $y$-coordinate.
Putting $\frac{\partial}{\partial y}=0$, and
redefining $\Psi\rightarrow e^{-i\int A_1\sigma^3 dx}\Psi$, we
have
\begin{equation}
H=2\int dx\left|\left(\partial_x
- iJ_x
\mp\frac{1}{2}(\sigma^3-Q^3)A_2\right)\Psi\right|^2
\mp 4 \pi T\,,
\label{redef}
\end{equation}
where 
$J_x = 
\frac1{2i}\Psi^\dagger\,
\partial_x ^{\hspace{-3.0mm}\raisebox{1.2mm}{$\leftrightarrow$}}
\Psi$.
$T$ is the one dimensional boundary term which is reduced 
through the Stokes theorem
from  $T_{[2]}$ defined in Eq.~(\ref{charg}),
\begin{equation}
4 \pi T = 
 (Q^3(x) -w) \times  A_2 (x) |_{x=-\infty}
- (Q^3(x) -w) \times  A_2 (x) |_{x=+\infty}\,.
\label{T}
\end{equation}
Solving the Gauss's law constraint, 
we get the explicit expression for the gauge field $A_2$,
\begin{equation}
A_2(x)=\frac{1}{2\kappa}
\int_{-\infty}^\infty K(x-y)(Q^3(y)+\rho_e)dy,
\label{solv}
\end{equation} 
where $K(x)$ is the one-dimensional kernel which 
solves the Eq.~(5): 
\begin{equation}
K(x) = \epsilon(x) + \beta.
\label{K}
\end{equation}
$\epsilon(x)$ 
is an odd-step function 
which is $\pm1$
depending on the signature of $x$,
and the constant $\beta$ is to be fixed by a boundary condition.
(See the discussion at the end of the next section.)

Now the  above Hamiltonian is  written in terms of 
matter field only and it gives the self-dual spin
system modified by a non-local interactions.  To see this 
let us expand the Hamiltonian explicitly which yields
\begin{equation}
H=2\int dx \left[ \left| \left(\partial_x 
-iJ_x
\right)\Psi \right|^2
\pm \frac{1}{2\kappa}(Q^3+\rho_e)(Q^3-w) +
h(x)\right]\,,
\label{1D_ham}
\end{equation}
where $h(x)$ is a non-local interaction term,  
$$h(x)=
\frac{1}{16\kappa^2}
(1-(Q^3)^2)\left|\int K(x-y)(Q^3(y)+\rho_e)dy\right|^2.$$
Modulo the nonlocal term $h(x)$, 
the above Hamiltonian is  exactly the same with
that of ILLE in the external magnetic field proportional to
$\rho_e -w$  in 1+1 dimension. 
The four parameters, $\kappa$,
$\rho_e$, $w$, and $\beta$  
control the spin direction
of the ground state,
and  the non-local term  $h(x)$ supplementing
the anisotropic potential energy plays an important role for
various solitons and domain walls
as we shall see in the next section.\\

\section{Soliton and Domain Wall Solutions}
The Hamiltonian in consideration  
is positive semi-definite and the
lower bound is saturated by the 
non-local Bogomol'nyi equation:
\begin{equation}
\left(\partial_x - iJ_x -\frac{1}{4\kappa}(\sigma^3-Q^3)
\int K(x-y)(Q^3(y)+\rho_e)dy\right)\Psi=0.\label{restr}
\end{equation}
For definiteness, we concentrate on the upper sign 
(self-dual case) in the 
Hamiltonian Eq.~(\ref{redef})  from now on without 
loss  of generality. 
(The lower sign (anti-self dual case) can be obtained
from the upper sign by switching
 $\kappa$ to $-\kappa$.) 
To solve the above equation, 
we assume the real ansatz for the wavefunction,
\begin{equation}
\Psi_1=\displaystyle\frac{1}{\sqrt{1+\rho}},
\quad \Psi_2= \sqrt {\rho \over {1+\rho}},
\end{equation}
similar to the complex projective coordinate.
Then, the current $J_x$ vanishes and $Q^3$ is expressed as
\begin{equation}
Q^3=\Psi^\dagger \sigma^3\Psi=\frac{1-\rho}{1+\rho}.
\end{equation}
Putting the above together, we find that the two component 
equation reduces to a single one
\begin{equation}
\left(\log\rho\right)'= - \frac{1}{\kappa}
\int_{-\infty}^\infty  K(x-y)(Q^3(y)+\rho_e)dy,
\end{equation}
which after differentiation on both sides yields
\begin{equation}
\left(\log\rho\right)''= 
- \displaystyle\frac{2}{\kappa}\left[\Big(\rho_e+
\frac{1-\rho}{1+\rho}\Big)\right].
\end{equation}

\begin{figure}[bth]
\begin{center}  
\epsfig{figure=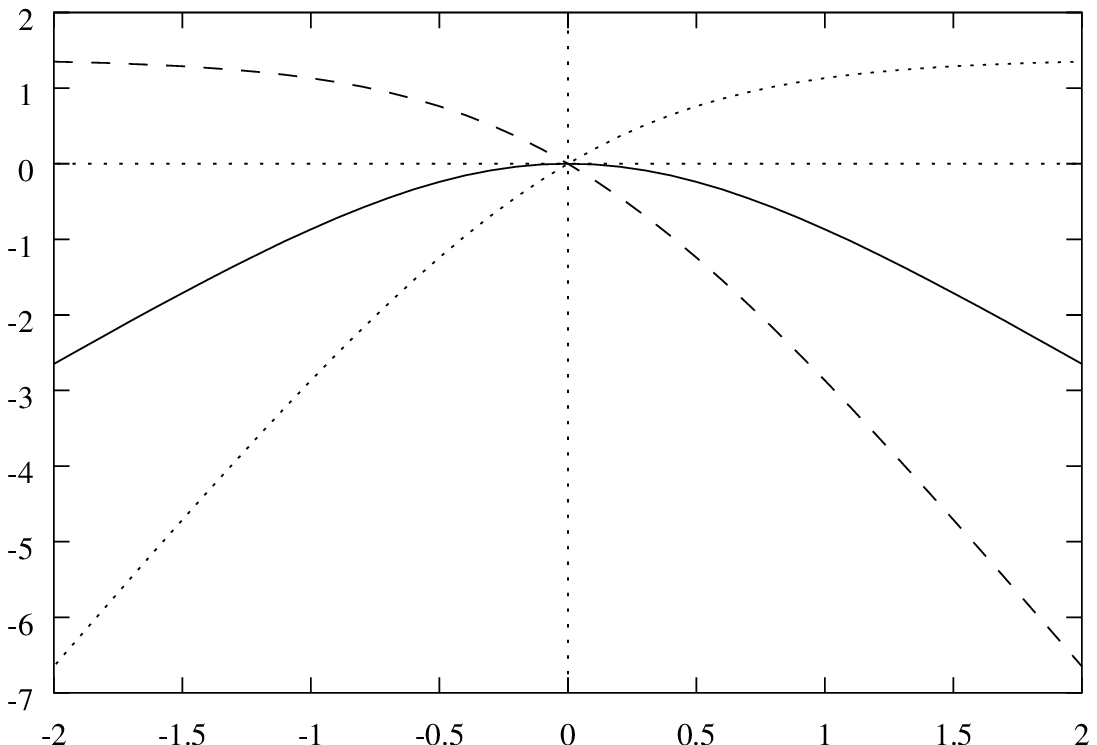,height=7cm}
\end{center}
\caption{The effective potential: $W(\phi)$ {\it v.s.} $\phi$ when
$\kappa>0$. The solid curve stands for a solution when $\rho_e=0$,
dotted
curve for $\rho_e=1$, and the dashed curve for $\rho_e=-1$.}
\label{fig-pot}
\end{figure}

\begin{figure}[bth]
\begin{center}
\epsfig{figure=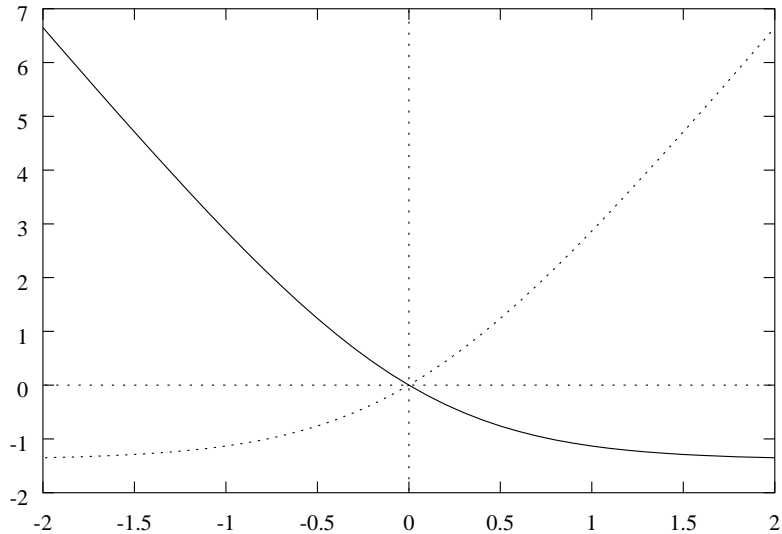,height=7cm}
\end{center}   
\caption{The effective potential: $W(\phi)$ {\it v.s.} $\phi$ when
$\kappa<0$. The solid curve stands for a solution when $\rho_e=1$ and
the dotted curve stands for $\rho_e=-1$}
\label{fig-pot2}
\end{figure}

This equation coincides precisely with the dimensionally reduced
vortex equation of Ref. \cite{ohpark,ohkim}.
To interpret this equation as the one dimensional Newtonian equation, 
we put 
$ \phi = { \ln \rho \over 2}$,
rescale the coordinate be $x \to x \sqrt{ |\kappa| } $,
and prepare the equation of motion in the form,
\begin{equation}
\phi'' = - {d W(\phi)  \over d \phi}, 
\quad 
\phi' = - \int_{-\infty}^\infty dy  K(x-y) (Q^3(y) + \rho_e), 
\label{phi}
\end{equation}
where $ W(\phi) =\mbox{sign}(\kappa)(
 - \ln \cosh \phi + \rho_e \phi)$,
whose form is depicted in Fig.~\ref{fig-pot} for $\kappa >0$ and in 
Fig.~\ref{fig-pot2} for $\kappa <0$.
We may view $W (\phi)$ as an effective potential
if we regard $x$ as ``time'' and 
$\phi$ as the position of the hypothetical ``particle" with unit mass.

Let us consider the solution for the case $\kappa >0$ first.
The effective potential $W(\phi)$ has a bump for 
$-1 < \rho_e < 1$. Therefore, 
if a soliton solution exists, then we expect that  
the soliton solution starts from $\phi = - \infty$ 
(``far left") at $x = - \infty$
and stops at the top of the hill as $x \to \infty$. 
This gives the boundary conditions for $\phi$,
\begin{equation}
\phi(x= - \infty) = -\infty\,,
\quad
\phi'(x = \infty) = 0.
\end{equation}
It is to be noted that the ``initial velocity" 
$\phi'( x = - \infty)$ should be infinite
since the ``total energy" is conserved.

If one considers the case in which
the ``particle" moves from the ``far right",
then  one may change the condition for the starting point as 
$\phi (x = -\infty) = \infty$. 
On the other hand, 
the condition for the final velocity is the same,
$\phi'(x = \infty) =0$. 
One may equally define the ``anti-particle" boundary condition
by switching the sign of the ``time":
the condition at $x= \infty$ 
is replaced by  that at $x = - \infty$ and {\it vice versa}.
Note that with this definition,
the ``anti-particle" of the particle
moving from ``far left" and ``particle" moving from
the ``far right" to the top of the hill does not coincide.  
To distinguish this possibility, we will call 
the condition $\phi(-\infty) = -\infty$
as the ``north-pole" condition 
($Q^3 (-\infty) =1$) and 
the condition $\phi(-\infty) = \infty$
as the ``south-pole" condition
($Q^3 (-\infty) = -1$).

\begin{figure}[bth]
\begin{center}
\epsfig{figure=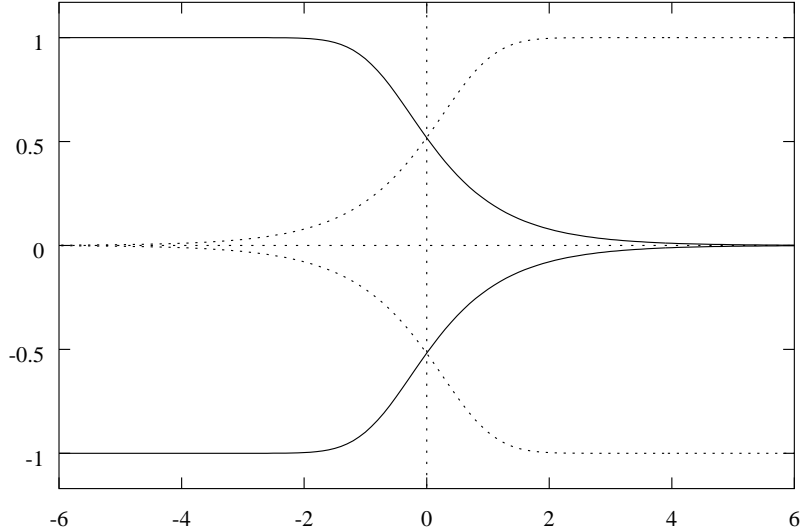,height=7cm}
\end{center}
\caption{Topological solitons: $Q^3(x)$ {\it v.s.} when $\rho_e=0$.
The solid curves stand for solitons and the dotted for an
anti-solitons.}
\label{tsol}
\end{figure}

\begin{figure}[bth]
\begin{center}
\epsfig{figure=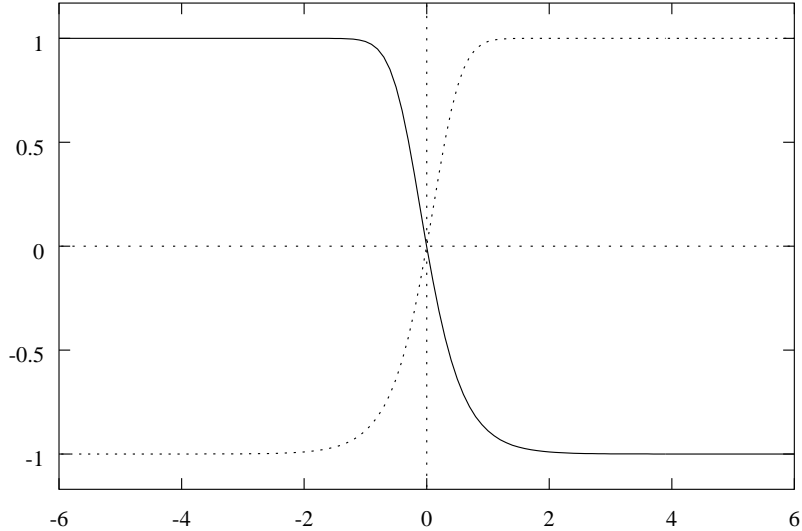,height=7cm}
\end{center}
\caption{North-pole solitons:
$Q^3(x)$ {\it v.s.} when
$\rho_e=1$ and
$\kappa>0$. The solid curve stands for a lump and the dotted for an
anti-lump.}
\label{npsol}
\end{figure}

\begin{figure}[bth]
\begin{center}
\epsfig{figure=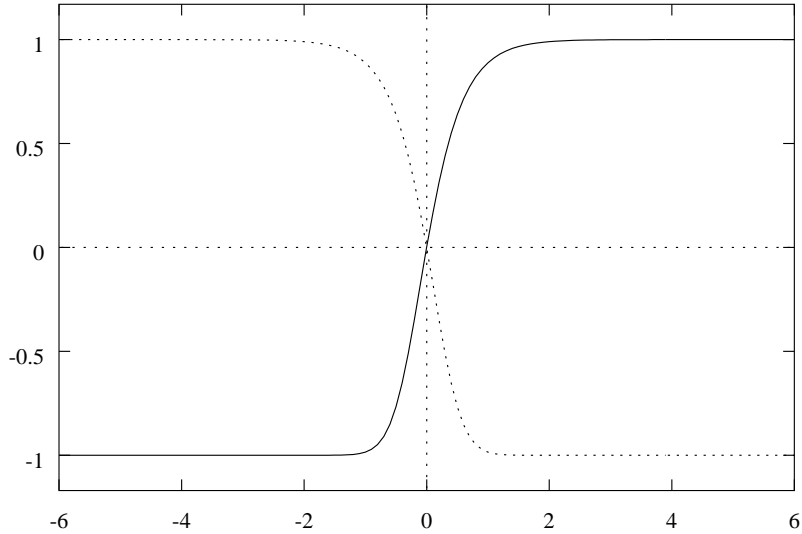,height=7cm}
\end{center}  
\caption{South-pole solitons: $Q^3(x)$ {\it v.s.} when $\rho_e=-1$ and
$\kappa>0$. The solid curve stands for a lump and the dotted for an
anti-lump.}
\label{spsol}
\end{figure}

We present  a numerical solution in Fig.~\ref{tsol} for $\rho_e =0$.
If we define a topological number,
\begin{equation}
q = {1 \over 2}( Q^3(x= -\infty) - Q^3(x= \infty))\,,
\end{equation}
then this soliton solution has 
the  topological number
$q=1/2$ for the  north-pole condition 
since $\phi$ goes from $\infty$ to $0$ (or $Q^3$ from $1$ to $0$).
And $q=-1/2$ for the  south-pole  condition since 
$\phi$ goes from $\infty $ to $0$ (or $Q^3$ from $-1$ to $0$).
The anti-soliton has the negative of the topological  number
of the soliton.
There are similar solutions for other values of $|\rho_e|<1$.
The topological number 
changes smoothly from $\pm {1 \over 2}$  up to $0$ or $\pm 1$
as $\rho_e$ varies: $q = {1 \over 2} (\pm 1 + \rho_e)$.
When the limiting value $\rho_e=  1 (-1)$ reaches, 
we have a solution corresponding  to a lump (anti-lump)
which has  $q= 1 (-1)$. 
(See Fig.~\ref{npsol} and Fig.~\ref{spsol}.)
We remark that the boundary condition for
the south-pole lump is the same as that 
for the north-pole  anti-lump.

\begin{figure}[tbh]
\begin{center}
\epsfig{figure=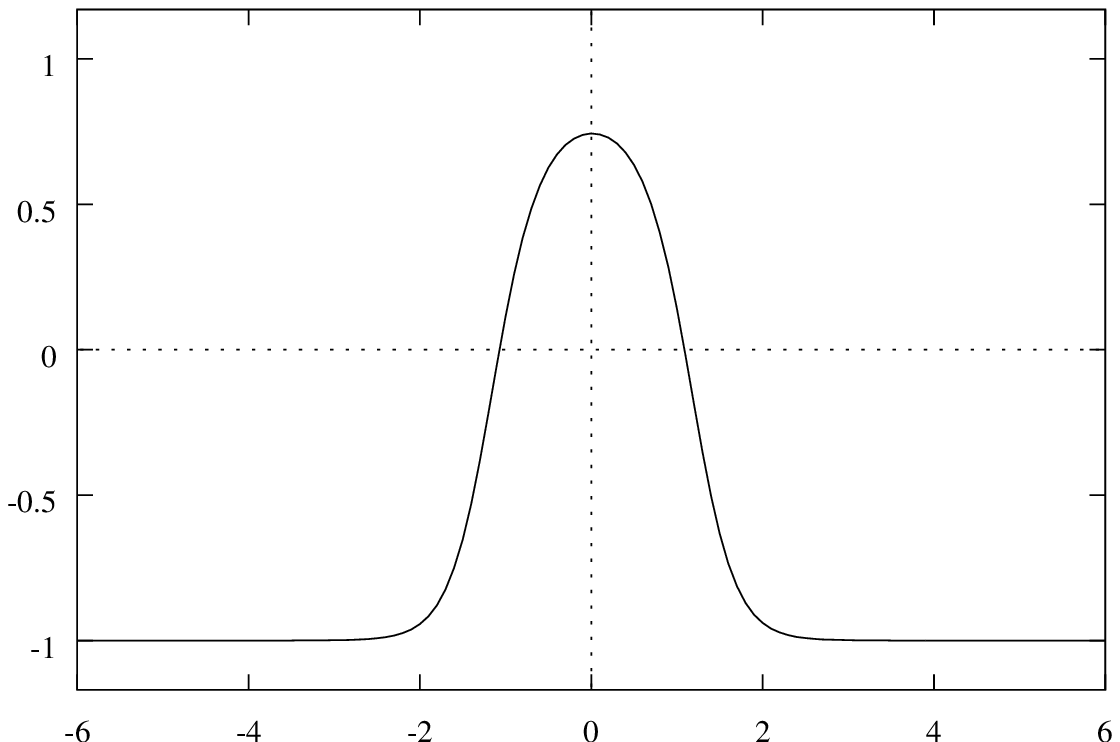,height=7cm}
\end{center}  
\caption{Non-topological solitons: $Q^3(x)$ {\it v.s.} when $\rho_e=1$
and $\kappa<0$.}
\label{ntsol}
\end{figure}

\begin{figure}[bth]
\begin{center}
\epsfig{figure=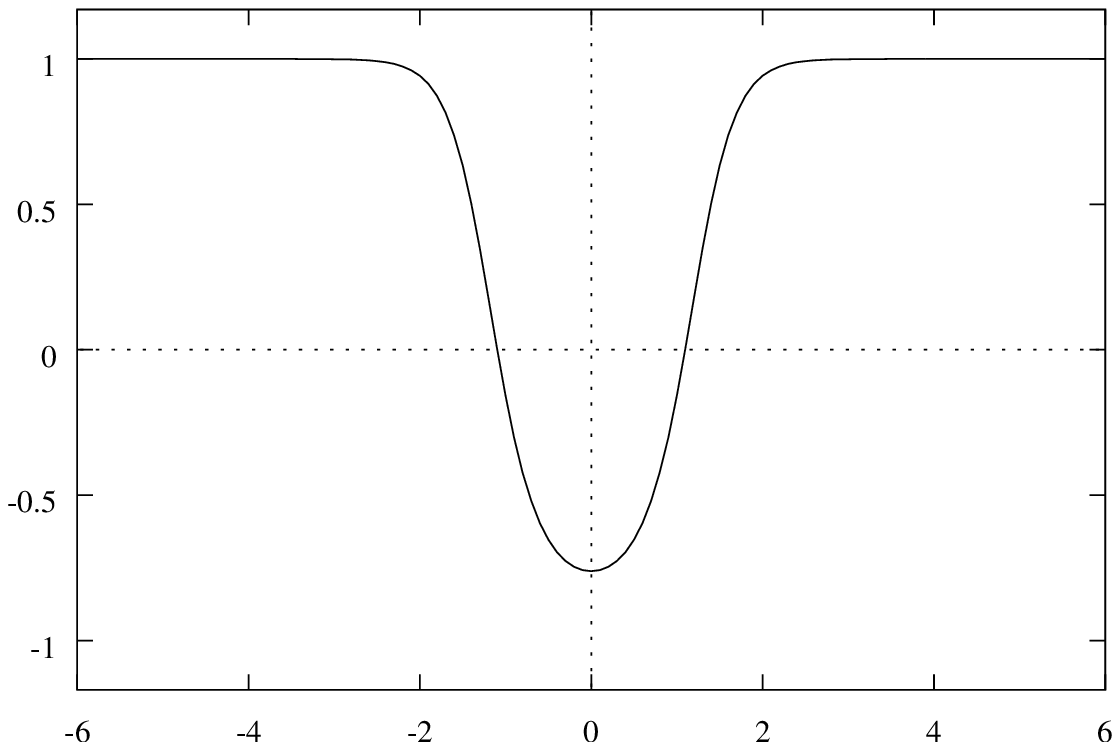, height=7cm}
\end{center}
\caption{Dark solitons: $Q^3(x)$ {\it v.s.} when $\rho_e=-1$ and
$\kappa<0$.}  
\label{dark}  
\end{figure}

When $\kappa <0$, we do not have a topological soliton solution 
as seen in the $\kappa >0$
case because  the effective potential $W(\phi)$ does 
not have a bump. (Fig.~\ref{fig-pot2}.)
Instead, we have a non-topological soliton
 with $\rho_e =1$ in which  the ``particle"  
starts from $\phi = +\infty$ (south-pole condition), reaches a 
turning point where it stops, changes the direction,
 and finally comes back 
to the original point $\phi = +\infty$ 
($Q^3$ starts from $-1$ and ends up with $-1$). (Fig.~\ref{ntsol}.)
For the other case $\rho_e =-1$, we have a dark soliton
in which the ``particle"  starts from the ``position"
$\phi = -\infty$ (north-pole condition),  reaches a 
point where it stops, changes the direction,  and finally comes back 
to the original ``position"  $\phi = -\infty$.
(Fig.~\ref{dark}.)

Let us turn to  the extra boundary term $T$ of  Eq.~(\ref{T}). 
Since $T$  adds to the soliton energy,
$T$ has to be  finite 
and be related with a  topological quantity 
if the solitons are  to be dynamically stable.
We note  that $T$ contains the value of  $A_2(x)$ at the boundary, 
$x = \pm \infty$ and that $A_2(x)$ is  proportional to 
the ``particle velocity", $\phi'(x)$:
$\phi'(x) = -2 A_2(x)$. 
Now, in the case of
topological solitons 
( $\kappa >0$ and $|\rho_e| \le 1$), 
$\phi'$ cannot be finite 
simultaneously at both ends ($x=\pm \infty$).
This means that 
$T$ is inevitably $\infty$ unless there occur 
some delicate  cancellations. 
The only way to make $T$ finite 
for the topological solution is to make $T$  vanish
by choosing the free parameter $w$ appropriately.
For the ``particle" case, $A_2(x = \infty) =0$ and 
we therefore,  must choose $w = Q^3(x = -\infty)$. 
For the ``anti-particle" case, $A_2(x = -\infty) =0$,
and $w = Q^3(x = \infty)$. 

For the non-topological soliton
($\kappa<0$ and $|\rho_e| =1$),
we see that the $\phi'$ has  finite values
at both ends ($x=\pm \infty$) with
$\phi'(- \infty) = -\phi'(\infty)$ 
($Q^3(\infty) = Q^3(-\infty))$,
and therefore, $T$ is finite with any value of 
$\vert\omega\vert\leq 1$.
The Hamiltonian is bounded by $- 4 \pi T$ whose value
is positive definite when $w \ne Q^3 (\infty)$.
We can certainly make $T$ vanish by choosing  
$w = Q^3(\infty)$,
whose dynamical stability betrays
the non-topological nature of the solution.
Nevertheless, the non-topological soliton can exist with 
the value of $w$ which turns out to be the maximal value for $Q^3$  in
Fig.~\ref{ntsol} 
(minimal value for dark soliton in Fig.~\ref{dark}).

One may suspect that there is a solution 
for $\kappa >0$ and $|\rho_e| <1$, 
where a ``particle"  starts to move at $\phi = - \infty$ and 
jump over the top of the potential energy and goes to the other side,
$\phi = + \infty$ (see Fig.~\ref{fig-pot}).  It turns out that 
this configuration is not energetically stable and does not 
exist.  This is because 
$\phi'(-\infty) = \phi'(\infty)$
and  $Q^3(-\infty) \ne Q^3(\infty)$. So $Q^3(-\infty)$ 
and $Q^3(\infty)$ have to  be a pair 
$(1, -1)$ or $(-1, 1)$. These values, however, make  
$T$ of   Eq.~(\ref{T}) infinite.
This is the reason why we do not have 
the lump soliton for $|\rho_e| < 1$ when $\kappa >0$.

In addition to the 
parameter $w$  in the Hamiltonian, 
Eq.~(\ref{redef}) or Eq.~(\ref{1D_ham}), 
the parameter $\beta$ 
(in the  one-dimensional kernel $K(x)$ 
in Eq.~(\ref{phi})) is to be fixed properly 
for the soliton solution to exist.
It is fixed  by the boundary value of $A_2(x)$.
Noting that $A_2(x)$ is proportional to
$\phi'(x)$, the condition for $A_2$ is obtained by 
the ``velocity" at the $\pm$  spatial infinity.
According to Eq.~(\ref{phi}), we have  
\begin{equation}
\phi'(\infty) = -(1 + \beta) \int_{- \infty}^{\infty}
dy (Q^3(y) + \rho_e)\,,\quad
\phi'(-\infty) = -(-1 + \beta) \int_{- \infty}^{\infty}
dy (Q^3(y) + \rho_e)\,.
\label{phi_prime}
\end{equation}
Since the integration of $Q^3 + \rho_e$ 
over the whole space need not be zero,
we have to choose $\beta =-1$ for the ``particle" 
to stop at the top of the hill
and $\beta = 1$ for the  ``anti-particle".
Therefore, our model allows  for $\kappa >0$
either the  soliton or the anti-soliton but not the both. 
For the non-topological case ($\kappa <0$), $\beta$ is 
to be fixed to be null.\\

\section{Conclusion}
We carried out a detailed study on the nonlocal self-dual
non-relativistic $CP(1)$ system in 1+1 dimension which results from the
dimensional reduction of the Abelian 
self-dual Chern-Simons $CP(1)$ system in 2+1 dimension. 
We found that the first order non-local Bogomol'nyi equation
yields a second order local equation which offers various soliton
solutions.

Our model Hamiltonian, Eq. (\ref{redef}) or Eq. (\ref{1D_ham})
 possesses four parameters $\kappa$, $\rho_e$, $w$, and 
$\beta$, which determine the properties of the soliton and domain wall.
The Bogomol'nyi solutions exist only for the specific choice of the 
parameters.  The symmetry of the Hamiltonian 
under the change of these parameters gives a useful 
information on the nature of the solutions. 
For example, self-dual solution with given $\kappa$ 
corresponds to the anti-self dual solution  
with $-\kappa$. Of course, this does not mean that 
there is one-to-one correspondence between a  
solution with $\kappa$  and another solution with 
$-\kappa$ for the self-dual case
or for the anti-self dual case respectively.

The system also possesses the interchnage symmetry between 
the north and the south-pole, $Q^3 \to  - Q^3$.
This symmetry is achieved by 
$\Psi  \to \sigma^1 \Psi$
and the simmultaneous change 
of the signs of two parameters
$(\rho_e, w)$.
Therefore, a self-dual (anti-self dual) 
solution with the north-pole condition
and $(\rho_e, w)$ guarantees the existence of 
a self-dual solution (anti-self dual)
solution with the south-pole solution 
and $(-\rho_e, -w)$.
 
In addition, parity symmetry (or ``time-reversal")
defined by $x\rightarrow -x,~ \Psi\rightarrow \pm\Psi$ 
is broken explicitly in the case when we have
non-vanishing values of the parameter $\beta$. 
But the Hamiltonian can still be rendered invariant
if the parity transformation is accompanied by the
transformation $\beta\rightarrow -\beta$.
This means that if there is a soliton (``particle") solution with a
given value of $\beta$, there exist an anti-soliton (``anti-particle")
solution corresponding to $-\beta$ which is parity inverted
(``time reversed") or {\it vice versa}.


Finally, we remark that 
the non-local interaction originates from the gauge field
$A_2$ which is eliminated through the Gauss's law constraint,
 but is responsible for the non-gauge type covariant 
derivative in the Hamiltonian, Eq. (\ref{redef}). 
This covariant derivative can be 
replaced by the ordinary derivative if we 
transform the wavefunction $\Psi
\rightarrow \Psi^\prime(\Psi)$ in a suitable way. 
Then, with $T=0$, the Hamiltonian becomes free in terms of $\Psi^\prime$. 
But this non-linear and non-local transformation 
does not preserve the symplectic structure of the phase
space of $\Psi$. (Qualitatively, the same is true in 
the NLSE case \cite{ohrim}).
It would be interesting to investigate further 
the nature of such a non-local and non-linear transformation
which renders the Hamiltonian free.

\vspace{1.5cm}
\noindent
{\large\bf Acknowledgements} \\
\indent
We would like to thank Dr. K. Kimm for useful discussions.
This work is supported in part
by the Korea Science and Engineering Foundation
through the Center for Theoretical Physics  at
Seoul National University and the project number
(95-0702-04-01-3, 94-1400-04-01-3),
and  by the Ministry of Education through the
Research Institute for Basic Science  (BSRI/97-1419, 97-2434).\\

\end{document}